\documentclass[12pt]{article}
\usepackage{ctex}
\usepackage{amssymb,amsmath,color,amsmath,bm}
\usepackage[colorlinks,linkcolor=blue]{hyperref}
\usepackage{geometry}
\usepackage{tabularx}
\usepackage{bm}
\usepackage{authblk}
\usepackage[square, comma, sort&compress, numbers]{natbib}

	\title{\bf The $(+)$-extended twisted generalized Reed-Solomon code}	
		\author{\small Canze Zhu}
	\author{\small Qunying Liao}
	\affil[]{\small (College of Mathematical Science, Sichuan Normal University, Chengdu Sichuan, 610066)}
	\date{}
	\newtheorem{theorem}{Theorem}[section]
	\newtheorem{definition}{Definition}[section]
	\newtheorem{lemma}{Lemma}[section]
	
	\newtheorem{proposition}{Proposition}[section]
	\newtheorem{corollary}{Corollary}[section]
	\newtheorem{remark}{Remark}[section]

	\catcode`@=11 \@addtoreset{equation}{section} \catcode`@=12
\begin{document}
	\maketitle
	{\bf Abstract.}
	{\small In this paper, we give  a parity check  matrix for the $(+)$-extended twisted generalized Reed Solomon (in short, ETGRS) code, and then not only prove that it is MDS or NMDS, but also determine the weight distribution. Especially, based on Schur method, we show that the $(+)$-ETGRS code is not GRS or EGRS.
Furthermore,  we present a  sufficient and necessary condition for any punctured code of the $(+)$-ETGRS code to be self-orthogonal, and then construct several classes of 
self-dual $(+)$-TGRS codes and  almost self-dual $(+)$-ETGRS codes. 
	}\\
	
	{\bf Keywords.}	{\small $(+)$-extended twisted generalized Reed Solomon codes;  MDS codes; NMDS codes; Self-dual codes; Almost self-dual codes.}
	
	\section{Introduction}
	 
	 An $[n,k,d]$ linear code $\mathcal{C}$ over $\mathbb{F}_q$ is a $k$-dimensional subspace of $\mathbb{F}_q^n$ with minimum (Hamming) distance $d$ and length $n$. If the
	 parameters reach the Singleton bound, namely, $d=n-k+1$, then $\mathcal{C}$  is  maximum distance separable (in short, MDS). If $d=n-k$, then
	  $\mathcal{C}$ is called almost MDS (in short, AMDS). In addition, $\mathcal{C}$ is said to be near MDS (in short, NMDS) if both $\mathcal{C}$ and $\mathcal{C}^{\perp}$ are AMDS. Since MDS codes and NMDS codes are very important in coding theory and applications \cite{7,17,28,35,43}, the study of MDS codes or NMDS codes, including weight distributions, constructions, equivalence, self-orthogonal, (almost) self-dual property, 
	   and so on, has attracted a lot of attention \cite{1,11,12,13,14,15,20,21,33,34,37}. Especially, generalized Reed-Solomon (in short, GRS) codes are a class of MDS codes. A lot of self-dual or almost self-dual MDS codes are constructed  based on GRS codes \cite{FX20,HZ21,10,19,25,39,40,LCD1,LCD2, LCD3}.

	In 2017,  inspired by the construction for twisted Gabidulin codes \cite{t26}, Beelen et al. firstly introduced  twisted Reed-Solomon (in short, TRS) codes, which is a generalization for Reed-Solomon codes, they also showed that TRS codes could be well decoded.  Different from GRS codes, they showed that a twisted generalized Reed-Solomon  (in short, TGRS) code is not necessary MDS and presented a  sufficient and necessary condition for a TGRS code to be MDS  \cite{2}. Especially, the authors showed that most of TGRS MDS codes are not GRS when the code rate  is less than one half \cite{3,26}. Later, by TGRS codes, Lavauzelle et al. presented an efficient key-recovery attack used in the McEliece cryptosystem \cite{24}. TGRS codes are also used to construct LCD MDS codes by their applications in cryptography \cite{16,26}. Recently, Huang et al. not only gave the parity check matrix for  the $(+)$-TGRS code, but also showed that it is MDS or NMDS. Furthermore, they presented a sufficient and necessary condition for the $(+)$-TGRS code to be self-dual, and then constructed several classes of self-dual MDS or NMDS codes \cite{t0}. More relative results about self-orthogonal  MDS or NMDS TGRS codes can be seen in \cite{ZC21,SD0,SD2}.

	In this paper, we focus on the $(+)$-extended twisted generalized Reed Solomon (in short, ETGRS) code. This paper is organized as follows. In section 2, some basic notations and results about linear codes are given. In section 3,  a parity check  matrix and the weight distribution for the $(+)$-ETGRS code are obtained, and then based on the Schur product method, we show that the $(+)$-ETGRS code is not GRS or EGRS. In section 4, a  sufficient and necessary condition for any punctured code of the $(+)$-ETGRS code to be self-orthogonal is presented, and then several classes of self-dual or almost self-dual  are constructed. In section 5, we conclude the whole paper.
	\section{Preliminaries}
	
	Throughout this paper, we fix some notations as follows for convenience.\\
	
	$\bullet$  $q$ is a power of  a prime.\\	
	
	$\bullet$  $\mathbb{F}_q$ is the finite field with $q$ elements, and $\mathbb{F}_q^{*}=\mathbb{F}_q\backslash\{0\}$.\\
	
	$\bullet$  $\mathbb{F}_q[x]$ is the polynomial ring over $\mathbb{F}_q$.\\
	
	$\bullet$  $k$ and $n$ are both positive integers with $2\le k<n$.\\

	$\bullet$  $\boldsymbol{1}=(1,\ldots,1)$, ~$\boldsymbol{0}=(0,\ldots,0)$.\\
	
	$\bullet$  For any $\boldsymbol{\alpha}=(\alpha_1,\ldots,\alpha_n)\in\mathbb{F}_q^n$, denote  $(\boldsymbol{\alpha},a)=(\alpha_1,\ldots,\alpha_n,a)~~(\forall a\in\mathbb{F}_{q})$,$$\boldsymbol{\alpha}^i=\begin{cases}
	(1,\ldots,1),&~~\text{if~}i=0;\\
	(\alpha_1^i,\ldots,\alpha_n^{i}),&~~\text{if~}i\in\mathbb{Z}^{+},
	\end{cases}~~~
	A_{\boldsymbol{\alpha}}=\{\alpha_i~|~i=1,\ldots,n\},\quad S_{\boldsymbol{\alpha}}=\sum\limits_{\alpha\in A_{\boldsymbol{\alpha}}}\alpha.$$
	
	
	In this section, we  review some basic notations and knowledge about GRS codes, EGRS codes, $(+)$-TGRS codes, $(+)$-ETGRS codes, Schur product, punctured codes, self-orthogonal codes,   NMDS codes and the subset sum problem, respectively.
	\subsection{GRS, EGRS, TGRS and ETGRS codes}
	\subsubsection{The GRS and EGRS code}
	The definitions of the GRS code and the EGRS code are given in the following, respectively.
		\begin{definition}[\cite{17}]\label{d0} 
		Let $\boldsymbol{\alpha}=(\alpha_1,\ldots,\alpha_n)\in\mathbb{F}_q^n$ with $\alpha_i\neq \alpha_j$ $(i\neq j)$ and
		$\boldsymbol{v} = (v_1,\ldots,v_{n})\in (\mathbb{F}_q^{*})^{n}$. Then
		the GRS code is defined as
		\begin{align*}
		\mathcal{GRS}_{k,n}(\boldsymbol{\alpha},\boldsymbol{v})=\{(v_1f({\alpha_{1}}),\ldots,v_nf(\alpha_{n}))~|~f(x)\in\mathbb{F}_q[x],~\deg f(x)\le k-1\}.
		\end{align*}	
		The EGRS code is defined as
		\begin{align*}
		\mathcal{GRS}_{k,n}(\boldsymbol{\alpha},\boldsymbol{v},\infty)=\{(v_1f({\alpha_{1}}),\ldots,v_nf(\alpha_{n}),f_{k-1})~|~f(x)\in\mathbb{F}_q[x],~\deg f(x)\le k-1\}
		\end{align*}	
		where  $f_{k-1}$ is the coefficient of $x^{k-1}$ in $f(x)$.
		
		If $\boldsymbol{v}=\boldsymbol{1}$, then $\mathcal{GRS}_{k,n}(\boldsymbol{\alpha},\boldsymbol{1})$ and $\mathcal{GRS}_{k,n}(\boldsymbol{\alpha},\boldsymbol{1},\infty)$ are the RS code and the ERS code, respectively.
	\end{definition}

	The dual codes of the GRS code and the EGRS code are given in the following, respectively.
	\begin{lemma}[\cite{19}]\label{drs}
		Let $\boldsymbol{u}=(u_1,\ldots,u_n)$ with $u_j=-\prod\limits_{i=1,i\neq j}^{n}(\alpha_j-\alpha_i)^{-1}$, then
		$$\big(\mathcal{GRS}_{k,n}(\boldsymbol{\alpha},\boldsymbol{1})\big)^{\perp}=\mathcal{GRS}_{n-k,n}(\boldsymbol{\alpha},\boldsymbol{u})$$ and
		$$\big(\mathcal{GRS}_{k,n}(\boldsymbol{\alpha},\boldsymbol{1},\infty)\big)^{\perp}=\mathcal{GRS}_{n+1-k,n}(\boldsymbol{\alpha},\boldsymbol{u},\infty).$$
	\end{lemma}
	\subsubsection{The $(+)$-TGRS and $(+)$-ETGRS code}
	\begin{definition}[\cite{2}]\label{d23} 
		Let $t$, $h$ and $k$ be positive integers with $0\le h<k\le q$ and $\eta\in\mathbb{F}_q^{*}$. Define the set of $(k,t,h,\eta)$-twisted polynomial as 
		\begin{align*}
			\mathcal{V}_{k,t,h,\eta}=\Big\{f(x)=\sum\limits_{i=0}^{k-1}a_ix^i+\eta a_hx^{k-1+t}~|~a_i\in\mathbb{F}_q~(i=0,\ldots,k-1)\Big\},
		\end{align*}
		which is a $k$-dimensional $\mathbb{F}_q$-linear subspace. $h$ and $t$ are the hook and the twist, respectively.
	\end{definition}

	From the twisted polynomials linear  space $\mathcal{V}_{k,1,k-1,\eta}$, the definitions of the $(+)$-TGRS code and the $(+)$-ETGRS code are given in the following, respectively.
	\begin{definition}[\cite{2}]\label{d1} 
	Let $\eta\in\mathbb{F}_q^{*}$, $\boldsymbol{\alpha}=(\alpha_1,\ldots,\alpha_n)\!\in\!\mathbb{F}_q^n$ with $\alpha_i\neq \alpha_j$ $(i\neq j)$ and
		$\boldsymbol{v} = (v_1,\ldots,v_{n})\in (\mathbb{F}_q^{*})^{n}$. Then
		the $(+)$-TGRS code is defined as
		\begin{align*}
		\mathcal{C}_{k,n}(\boldsymbol{\alpha},\boldsymbol{v},\eta)=\{(v_1f({\alpha_{1}}),\ldots,v_nf(\alpha_{n}))~|~f(x)\in  \mathcal{V}_{k,1,k-1,\eta}\}.
		\end{align*}	
		The $(+)$-ETGRS code is defined as
		\begin{align*}
			\mathcal{C}_{k,n}(\boldsymbol{\alpha},\boldsymbol{v},\eta,\infty)=\{(v_1f({\alpha_{1}}),\ldots,v_nf(\alpha_{n}),f_{k-1})~|~f(x)\in  \mathcal{V}_{k,1,k-1,\eta}\},
		\end{align*}	
		where  $f_{k-1}$ is the coefficient of $x^{k-1}$ in $f(x)$.

	 If $\boldsymbol{v}=\boldsymbol{1}$, then $\mathcal{C}_{k,n}(\boldsymbol{\alpha},\boldsymbol{1},\eta)$ and $\mathcal{C}_{k,n}(\boldsymbol{\alpha},\boldsymbol{1},\eta,\infty)$ are the $(+)$-TRS code and the $(+)$-ETRS code, respectively.
	\end{definition}

	\begin{remark}
	By Definition \ref{d23}, it is easy to see that the generator matrix for $\mathcal{C}_{k,n}(\boldsymbol{\alpha},\boldsymbol{v},\eta,\infty)$ is
	\begin{align*}
	G_k=\left(\begin{matrix}
	&v_1~&v_2~&\ldots~&v_n&0~\\
	&v_1\alpha_1~&v_2\alpha_2~&\ldots~&v_n\alpha_{n}&0~\\
	&\vdots~&\vdots&~&\vdots&\vdots~\\
	&v_1\alpha_1^{k-2}~&v_2\alpha_{2}^{k-2}~&\cdots~&v_n\alpha_{n}^{k-2}&0~\\
	&v_1(\alpha_1^{k-1}+\eta\alpha_1^{k})~&v_2(\alpha_2^{k-1}+\eta\alpha_2^{k})~&\ldots~&v_n(\alpha_{n}^{k-1}+\eta\alpha_h^{k})&1~\\
	\end{matrix}\right).
	\end{align*}
	\end{remark}
	
\subsection{Some notations for linear codes}
	\subsubsection{The Schur product}
The Schur product  is defined as follows.
\begin{definition}\label{SP}
	For $\mathbf{x}=(x_1,\ldots,x_n ), \mathbf{y}= (y_1 ,\dots, y_n)\in \mathbb{F}_q^{n}$,  the Schur product  between $\mathbf{x}$ and $\mathbf{y}$ is
	defined as $$\mathbf{x}\star\mathbf{y}:=(x_1y_1,\ldots,x_ny_n).$$  The Schur product  of two $q$-ary codes $\mathcal{C}_1$ and $\mathcal{C}_2$ with length $n$ is
	defined as
	\begin{align*}
	\mathcal{C}_1\star\mathcal{C}_2=\langle \mathbf{c}_1 \star\mathbf{c}_2~|~\mathbf{c}_1\in\mathcal{C}_1,\mathbf{c}_2\in\mathcal{C}_2\rangle.
	\end{align*}
	Especially, for a code $\mathcal{C}$, we call $\mathcal{C}^2:=\mathcal{C}\star\mathcal{C}$ the Schur square of $\mathcal{C}$.
\end{definition} 
\begin{remark}\label{r1}For any linear codes  $\mathcal{C}_1$ and $\mathcal{C}_2$, if 
	$\mathcal{C}_1=\langle \boldsymbol{v}_1,\ldots,\boldsymbol{v}_{k_1}\rangle$ and   $\mathcal{C}_2=\langle \boldsymbol{w}_1,\ldots,\boldsymbol{w}_{k_2}\rangle$ with $\boldsymbol{v}_i,\boldsymbol{w}_j\in\mathbb{F}_q^{n}~(i=1,\ldots,k_1,~j=1,\ldots,k_2)$, then 
	\begin{align}\label{S1}
	\mathcal{C}_1\star\mathcal{C}_2=\langle \boldsymbol{v}_i\star\boldsymbol{w}_j~(i=1,\ldots,k_1,~j=1,\ldots,k_2)\rangle, 
	\end{align}
\end{remark}

By the definitions of the GRS code and the EGRS code, Lemma \ref{drs} and Remark \ref{r1}, we have the following proposition about the Schur square of GRS (EGRS) code.
\begin{proposition}\label{pr}	Let $\boldsymbol{u}=(u_1,\ldots,u_n)$ with $u_j=-\prod\limits_{i=1,i\neq j}^{n}(\alpha_j-\alpha_i)$ $(j=1,\ldots,n)$.

	$(1)$ If $k\le\frac{n+1}{2}$, then {\small \begin{align*}
	\mathcal{GRS}^{2}_{k,n+1}(\boldsymbol{\alpha},\boldsymbol{1})=\mathcal{GRS}_{2k-1,n+1}(\boldsymbol{\alpha},\boldsymbol{1}) \text{~~~and~~~}\mathcal{GRS}_{k,n}^{2}(\boldsymbol{\alpha},\boldsymbol{1},\infty)=\mathcal{GRS}_{2k-1,n}(\boldsymbol{\alpha},\boldsymbol{1},\infty).
	\end{align*}}
		
	$(2)$ If $n+1\ge  k\ge\frac{n}{2}+1$, then{\small\begin{align*}
	\big(\mathcal{GRS}_{k,n+1}^{\perp}(\boldsymbol{\alpha},\boldsymbol{1})\big)^2=\mathcal{GRS}_{2n-2k+1,n+1}(\boldsymbol{\alpha},\boldsymbol{u}^2)\text{~~and~~}\big(\mathcal{GRS}_{k,n}^{\perp}(\boldsymbol{\alpha},\boldsymbol{1},\infty)\big)^2=\mathcal{GRS}_{2n-2k+1,n}(\boldsymbol{\alpha},\boldsymbol{u}^2,\infty).
	\end{align*}}
\end{proposition}
	\subsubsection{The equivalence and punctured codes for linear codes}
	The definition of the equivalence for linear codes is given in the following.
	
	\begin{definition}
		Let $\mathcal{C}_1$ and $\mathcal{C}_2$ be linear codes over $\mathbb{F}_q$ with length $n$, and $S_n$ be the permutation group with order $n$. We say that $\mathcal{C}_1$ and $\mathcal{C}_2$ are equivalent if
		there is a permutation $\pi\in S_n$, and $\boldsymbol{v}=(v_1,\dots,v_n)\in(\mathbb{F}_q^*)^{n}$ such that $\mathcal{C}_2=\Phi_{\pi,\boldsymbol{v}}(\mathcal{C}_1)$, where
		\begin{align*}
		\Phi_{\pi,\boldsymbol{v}}:\mathbb{F}_q^n\to\mathbb{F}_q^n,\quad(c_1,\ldots,c_n)\mapsto (v_1c_{\pi(1)},\ldots,v_nc_{\pi(n)}).
		\end{align*}
	\end{definition}
	\begin{remark}\label{es}
		It is easy to see that   $\mathcal{C}_1^2$ and $\mathcal{C}_2^2$ are equivalent when $\mathcal{C}_1$ and $\mathcal{C}_2$ are equivalent. 
	\end{remark}

The definition of the punctured code is given in the following.
\begin{definition}
	For any positive integers $m$ and $n$ with $m\le n$, let $\mathcal{C}$ be a linear code over $\mathbb{F}_q$ with length $n$, and $I=\{i_1,\ldots,i_m\}\subseteq \{1,\ldots,n\}$. The punctured code for  $\mathcal{C}$ over $I$ is defined as
	\begin{align*}
	\mathcal{C}_{I}=\{(c_{i_1},\ldots,c_{i_m})~|~(c_1,\ldots,c_n)\in\mathcal{C}\}. 
	\end{align*}
\end{definition}  
\begin{remark}\label{r0}
	$(1)$ If $\boldsymbol{v}_1,\boldsymbol{v}_2\in(\mathbb{F}_{q}^{*})^n$, $\boldsymbol{\alpha}_1,\boldsymbol{\alpha}_2\in\mathbb{F}_q^{n}$ with $A_{\boldsymbol{\alpha}_1}=A_{\boldsymbol{\alpha}_2}$, then $\mathcal{C}_{k,n}(\boldsymbol{\alpha}_1,\boldsymbol{v}_1,\eta,\infty)$ and $\mathcal{C}_{k,n}(\boldsymbol{\alpha}_2,\boldsymbol{v}_2,\eta,\infty)$ are equivalent.
	
	 $(2)$ If $A_{\boldsymbol{\alpha}}=\mathbb{F}_q$, we denote $\mathcal{C}_{k,n}(\mathbb{F}_q,\boldsymbol{v},\eta,\infty)=\mathcal{C}_{k,n}(\boldsymbol{\alpha},\boldsymbol{v},\eta,\infty)$. Obviously, any $(+)$-TGRS code or $(+)$-ETGRS code is equivalent to a punctured code of $\mathcal{C}_{k,n}(\mathbb{F}_q,\boldsymbol{v},\eta,\infty)$.
\end{remark}

	\subsubsection{Self-orthogonal linear codes}
	
	The notations about self-orthogonal, self-dual or almost self-dual codes are given in the following, respectively.
	
	For $\mathbf{a}=(a_1,\ldots,a_n)$, $\mathbf{b}=(b_1,\ldots,b_n)$ $\in\mathbb{F}_q^n$, the inner product is defined as $$\langle \mathbf{a},\mathbf{b}\rangle=\sum\limits_{i=1}^{n}a_ib_i.$$ And then 
	the dual code of $\mathcal{C}$ is defined as $$\mathcal{C}^{\perp}=\{\mathbf{c}^{'}\in\mathbb{F}_q^n~|~\langle\mathbf{c}^{'},\mathbf{c}\rangle=0, \text{for any}~\mathbf{c}\in\mathcal{C}\}.$$ Especially, if $\mathcal{C}\subseteq\mathcal{C}^{\perp}$, then $\mathcal{C}$ is self-orthogonal. Especially,	if $\mathcal{C}=\mathcal{C}^{\perp}$, then $\mathcal{C}$ is self-dual; if $\mathcal{C}$ is self-orthogonal with length $n$ odd and $\dim(\mathcal{C})=\frac{n-1}{2}$, then $\mathcal{C}$ is almost self-dual.   \\
	
	 Let $1_{\pi}$ be the identity in  $S_n$, a sufficient and necessary condition for  $\Phi_{1_{\pi},\boldsymbol{v}}(\mathcal{C}_I)$ to be self-orthogonal is presented in the following lemma.
	 	\begin{lemma}\label{lso}
	 		Let $n$ and $m$ be positive integers with $m\le n$,  $I=\{i_1,\ldots,i_m\}\subseteq \{1,\ldots,n\}$, and $\boldsymbol{v}=(v_1,\ldots,v_m)\in(\mathbb{F}_q^{*})^{m}$. Then for a $q$-ary linear code $\mathcal{C}$ with length $n$, $\Phi_{1_{\pi},\boldsymbol{v}}(\mathcal{C}_I)$ is self-orthogonal if and only if there is some
	 		$\mathbf{c}\in(\mathcal{C}^2)^{\perp}$ such that
	 		\begin{align*} \mathrm{Supp}(\mathbf{c})=I~~~\text{and}~~~c_{i_j}=v_j^2~(j=1,\ldots,m).
	 		\end{align*} 
	 \end{lemma}
 
	{\bf Proof}. By the definition of the self-orthogonal code,  $\Phi_{1_{\pi},\boldsymbol{v}}(\mathcal{C}_I)$ is self-orthogonal if and only if   
	\begin{align*} 
	\sum\limits_{j=1}^{m}v_{j}c_{1,i_j}v_jc_{2,i_j}=0~~\text{~for~any~~}\mathbf{c}_t=(c_{t,i_1},\ldots,c_{t,i_m})\in\mathcal{C}_{I}~(t=1,2),
	\end{align*}namely,
	\begin{align*}
	\sum\limits_{j=1}^{m}v_{j}^2(c_{1,i_j}c_{2,i_j})=0~~\text{~for~any~~}\mathbf{c}_t=(c_{t,i_1},\ldots,c_{t,i_m})\in\mathcal{C}_{I}~(t=1,2),
	\end{align*}
	equivalently, there is some $\mathbf{c}=(c_1,\ldots,c_n)\in(\mathcal{C}^2)^{\perp}$ such that \begin{align*} \mathrm{Supp}(\mathbf{c})=I~~~\text{and}~~~c_{i_j}=v_j^2~(j=1,\ldots,m).
	\end{align*}$\hfill\Box$
	\subsection{Near MDS codes and the subset sum problem over finite fields}
\subsubsection{Weight distributions of near MDS codes}

It is well-known that the weight distribution for the MDS code $[n,k,n−k+1]$ over $\mathbb{F}_q$ depends only on  the values of $n$, $k$ and $q$. But for the NMDS code $[n,k,n−k]$ over $\mathbb{F}_q$,  the weight distribution depends not only on the values of $n$, $k$ and $q$, but also on the  number of its minimum weight codewords, which can be seen in the following lemma.

\begin{lemma}(\cite{7})\label{NMDS}
	Let $\mathcal{C}$ be an $[n, k, n-k]$ NMDS code over $\mathbb{F}_q$ and $A_i$ $(i=0,1,\ldots,n)$ the number of codewords in $\mathcal{C}$
	with weight $i$. Then weight distributions of $\mathcal{C}$ and $\mathcal{C}^{\perp}$ are given by
	\begin{align}\label{ANMDS}
	A_{n-k+s}=\binom{n}{k-s}\sum\limits_{j=0}^{s-1}(-1)^j\binom{n-k+s}{j}(q^{s-j}-1)\!+\!(-1)^s\binom{k}{s}A_{n-k}\quad( s=1,\ldots,k),
	\end{align}
	and
	\begin{align}\label{DNMDS}
	A_{k+s}^{\perp}=\binom{n}{k+s}\sum\limits_{j=0}^{s-1}(-1)^j\binom{k+s}{j}(q^{s-j}-1)\!+\!(-1)^s\binom{k}{s}A_{k}^{\perp}\quad(s=1,\dots,n-k).
	\end{align}
	Furthermore, $$A_{n-k}=A_{k}^{\perp}.$$	
\end{lemma}
\begin{remark}
Let $\mathcal{C}$ be an $[n, k, n-k+1]$ MDS code, then $A_{n-k}=A_{k}^{\perp}=0$, the weight distributions for $\mathcal{C}$ and $\mathcal{C}^{\perp}$ are given in $(\ref{ANMDS})$ and $(\ref{DNMDS})$, respectively.
\end{remark}
\subsubsection{The subset sum problem over finite fields}
In the following, we give the notation and a lemma for the subset sum problem over finite fields, which are needed to determine the weight distribution of $\mathcal{C}_{k,n}(\boldsymbol{\alpha},\boldsymbol{v},\eta,\infty)$.

The subset sum problem is a well-known $\mathbf{NP}$-complete problem. Given $t\in\mathbb{Z}^{+}$, $b\in\mathbb{F}_q$ and $D\subseteq\mathbb{F}_q$ with $|D|\ge t$, let 
\begin{align*}
N(t,b,D)=\big\{\{x_1,\ldots,x_t\}\subseteq D\big|x_1+\cdots+x_t=b\big\},
\end{align*}
then $\#N(t,b,D)$ is the number of $t$-element subsets of $D$ whose sum is $b$. Determine the value of $\#N(t,b,D)$ is the subset sum problem. 

For $D=\mathbb{F}_q$ or $\mathbb{F}_q^{*}$, the value of $\#N(t,b,D)$ is given explicitly in the following  lemma.
\begin{lemma}(\cite{LD})\label{SSS}
	Let  $v(b)=\begin{cases}
	q-1,&\text{if~}b=0;\\
	-1,&\text{if~}b\neq 0.
	\end{cases}$ 	
	Then	
	
	\begin{align*}
	\#N(t,b,\mathbb{F}_{q}^{*})=
	\frac{1}{q}\binom{q-1}{t}+(-1)^{t+\lfloor\frac{t}{p}\rfloor}\frac{v(b)}{q}\binom{\frac{q}{p}-1}{\lfloor\frac{t}{p}\rfloor},
	\end{align*}	
	and
	\begin{align*}
	\#N(t,b,\mathbb{F}_{q})=\begin{cases}
	\frac{1}{q}\binom{q}{t},&~\text{if}~p\nmid t;\\
	\frac{1}{q}\binom{q}{t}+(-1)^{t+\frac{t}{p}}\frac{v(b)}{q}\binom{\frac{q}{p}}{\frac{t}{p}},&~\text{if}~p\mid t.
	\end{cases}
	\end{align*}
\end{lemma}
\begin{remark}\label{rsss}For $t\ge 2$, it is easy to see that $\#N(t,b,\mathbb{F}_{q}^{*})=0$ or $\#N(t,b,\mathbb{F}_{q})=0$ if and only if  $2\mid q$ and $(t,b)\in\{(2,0),~(q-2,0)\}$.
\end{remark}

\section{Properties for $(+)$-ETGRS codes} 
In this section, we give  a parity check  matrix for the $(+)$-ETGRS code, and then prove that $(+)$-ETGRS  is MDS or NMDS. Furthermore, we give a sufficient and necessary condition for a $(+)$-ETGRS code to be MDS or NMDS, and then determine the weight distribution based on the subset sum problem. Finally, by using Schur method, we show that it is not GRS or EGRS.

\subsection{ A parity check  matrix  for the $(+)$-ETGRS code }
 The following lemma is necessary to calculate  a parity check  matrix for the $(+)$-ETGRS code.
 \begin{lemma}\label{lu}
	For any $m\in\mathbb{N}$ and $A\subseteq\mathbb{F}_q$ with $|A|>2$, let 
	$$L_A(m)=\sum\limits_{\alpha\in A}\alpha^{m}\prod_{\beta\in\mathbb{F}_q\backslash A}(\alpha-\beta),$$ then
	\begin{align}\label{lq}
	L_A(m)=\begin{cases}
	0,\quad&\text{if}~m\le |A|-2;\\
	-1,\quad&\text{if}~m=|A|-1;\\
	-\sum\limits_{\alpha\in A}\alpha,&\text{if}~m=|A|.
	\end{cases}
	\end{align}
\end{lemma}

{\bf Proof.} For any $l\in\mathbb{Z}^{+}$, it is well-known that 
\begin{align}\label{lq1}
\sum\limits_{\gamma\in \mathbb{F}_q}\gamma^{l}=\begin{cases}
-1,\quad&\text{if}~(q-1)\mid l;\\
0,\quad&\text{otherwise}.
\end{cases}
\end{align}
Note that \begin{align*}
\prod_{\beta\in\mathbb{F}_q\backslash A}(\alpha-\beta)=&\alpha^{q-|A|}-\sum\limits_{\beta\in\mathbb{F}_q\backslash A}\beta\alpha^{q-|A|-1}+\cdots+(-1)^{q-|A|}\prod_{\beta\in\mathbb{F}_q\backslash A}\beta\\
=&\alpha^{q-|A|}+\sum\limits_{\gamma\in A}\gamma\alpha^{q-|A|-1}+\cdots+(-1)^{q-|A|}\prod_{\beta\in\mathbb{F}_q\backslash A}\beta,
\end{align*}
we have 
\begin{align}\label{lq2}	\begin{aligned}
L_A(m)=&\sum\limits_{\alpha\in A}\alpha^{m}\prod_{\beta\in\mathbb{F}_q\backslash A}(\alpha-\beta)\\
=&\sum\limits_{\alpha\in \mathbb{F}_{q}}\Big(\alpha^{q-|A|+m}+\sum\limits_{\gamma\in A}\gamma\alpha^{q-|A|-1+m}+\cdots+(-1)^{q-|A|+m}\prod_{\beta\in\mathbb{F}_q\backslash A}\beta\alpha^m\Big).
\end{aligned}\end{align}
Now by $(\ref{lq1})$-$(\ref{lq2})$, we obtain $(\ref{lq})$ directly. $\hfill\Box$\\

In the following theorem, we give a parity check  matrix of the $(+)$-ETGRS  code.
	\begin{theorem}\label{dt}Let $u_j=-\prod\limits_{i=1, i\neq j}^{n}(\alpha_j-\alpha_i)^{-1}$ $(j=1,\ldots,n)$, then 	
	$\mathcal{C}_{k,n}(\boldsymbol{\alpha},\boldsymbol{v},\eta,\infty)$ has  a parity check  matrix
	\begin{align*}
	H_{n+1-k}=\left(\begin{matrix}
	&\frac{u_1}{v_1}~&\frac{u_2}{v_2}~&\ldots~&\frac{u_n}{v_n}&0~\\
	&\frac{u_1}{v_1}\alpha_1~&\frac{u_2}{v_2}\alpha_2~&\ldots~&\frac{u_n}{v_n}\alpha_n&0~\\
	&\vdots~&\vdots&~&\vdots&\vdots~\\
	&\frac{u_1}{v_1}\alpha_1^{n-k-1}~&\frac{u_2}{v_2}\alpha_{2}^{n-k-1}~&\cdots~&\frac{u_n}{v_n}\alpha_{n}^{n-k-1}&\eta~\\
	&\frac{u_1}{v_1}\alpha_1^{n-k}~&\frac{u_2}{v_2}\alpha_{2}^{n-k}~&\cdots~&\frac{u_n}{v_n}\alpha_{n}^{n-k}&1+\eta S_{\boldsymbol{\alpha}}\\
	\end{matrix}\right).
	\end{align*}
	Namely, $$\mathcal{C}_{k,n}^{\perp}(\boldsymbol{\alpha},\boldsymbol{v},\eta,\infty)=\Big\{\Big(\frac{u_1}{v_1}g({\alpha_{1}}),\ldots,\frac{u_n}{v_n}g(\alpha_{n}),\eta g_{n-k-1}+(1+\eta S_{\boldsymbol{\alpha}})g_{n-k}\Big)~\Big|~\deg g(x)\le n-k\Big\},$$
	where  $g_{n-k-1}$ and $g_{n-k}$ are the coefficients of $x^{n-k-1}$ and $x^{n-k}$ in $g(x)$, respectively.
	\end{theorem}

{\bf Proof}. Let
 $$G_k=\left(\begin{matrix}\!\!&\boldsymbol{g}_0~\\
&\!\!\vdots~\\
&\!\!\boldsymbol{g}_{k-1}\end{matrix}\right)\quad \text{and}\quad 
H_{n+1-k}=\left(\begin{matrix}	&\!\!\boldsymbol{h}_0~\\
&\!\!\vdots~\\
&\!\!\boldsymbol{h}_{n-k}\end{matrix}\right).$$ Note that $\mathrm{Rank} (H_{n+1-k})=n+1-k$, thus it is enough to prove that 
\begin{align}\label{gh}
\langle\boldsymbol{g}_s,\boldsymbol{h}_l\rangle=0\quad(\forall s\in\{0,\ldots,k-1\},~l\in\{0,\ldots,n-k\}).
\end{align}
By  Lemma \ref{lu} and $$u_j=-\prod\limits_{i=1, i\neq j}^{n}(\alpha_j-\alpha_i)^{-1}=\prod\limits_{\beta\in\mathbb{F}_q\backslash A_{\boldsymbol{\alpha}}}(\alpha_j-\beta),$$ we can obtain $(\ref{gh})$ by the following four cases.

{\bf Case $1$}. For $s\in\{0,\ldots,k-2\}$ and $l\in\{0,\ldots,n-k\}$, we have $s+l\le n-2$, thus
\begin{align*}
		\langle\boldsymbol{g}_s,\boldsymbol{h}_l\rangle
	=&\sum\limits_{i=0}^{n}u_i\alpha_i^{s}\alpha_i^{l}=\sum\limits_{\alpha\in A_{\boldsymbol{\alpha}}}\alpha^{s+l}\prod_{\beta\in\mathbb{F}_q\backslash A_{\boldsymbol{\alpha}}}(\alpha-\beta)
	=0.
\end{align*}

{\bf Case $2$}. For $s=k-1$ and $l\in\{0,\ldots,n-k-2\}$, we have $k+l\le n-2$, thus
\begin{align*}
	\langle\boldsymbol{g}_s,\boldsymbol{h}_l\rangle
=&\sum\limits_{i=0}^{n}u_i(\alpha_i^{k-1}+\eta\alpha_i^{k})\alpha_i^{l}=\sum\limits_{\alpha\in A_{\boldsymbol{\alpha}}}(\alpha^{k-1+l}+\eta\alpha^{k+l})\prod_{\beta\in\mathbb{F}_q\backslash A_{\boldsymbol{\alpha}}}(\alpha-\beta)=0.
\end{align*}

{\bf Case $3$}. For $s=k-1$ and $l=n-k-1$, one has
\begin{align*}
	\langle\boldsymbol{g}_s,\boldsymbol{h}_l\rangle
=&\sum\limits_{i=0}^{n}u_i(\alpha_i^{k-1}+\eta\alpha_i^{k})\alpha_i^{n-k-1}+\eta\\
=&\sum\limits_{\alpha\in A_{\boldsymbol{\alpha}}}(\alpha^{n-2}+\eta\alpha^{n-1})\prod_{\beta\in\mathbb{F}_q\backslash A_{\boldsymbol{\alpha}}}(\alpha-\beta)+\eta=0.
\end{align*}

{\bf Case $4$}. For $s=k-1$ and $l=n-k$, one has
\begin{align*}
	\langle\boldsymbol{g}_s,\boldsymbol{h}_l\rangle
=&\sum\limits_{i=0}^{n}u_i(\alpha_i^{k-1}+\eta\alpha_i^{k})\alpha_i^{n-k}+(1+\eta S_{\boldsymbol{\alpha}})\\
=&\sum\limits_{\alpha\in A_{\boldsymbol{\alpha}}}(\alpha^{n-1}+\eta\alpha^{n})\prod_{\beta\in\mathbb{F}_q\backslash A_{\boldsymbol{\alpha}}}(\alpha-\beta)+(1+\eta S_{\boldsymbol{\alpha}})
=0.
\end{align*}

Now, by the above discussions, we complete the proof. $\hfill\Box$\\

\begin{corollary}\label{c1}
	$\mathcal{C}_{k,n}^{\perp}(\boldsymbol{\alpha},\boldsymbol{v},\eta,\infty)$ is MDS or AMDS.
\end{corollary}

{\bf Proof}. By Theorem \ref{dt}, for any codeword $\mathbf{c}\in\mathcal{C}_{k,n}^{\perp}(\boldsymbol{\alpha},\boldsymbol{v},\eta,\infty)\backslash\{\boldsymbol{0}\}$, there exists some $g(x)=\sum\limits_{i=0}^{n-k}g_ix^{i}\in\mathbb{F}_q[x]$ such that
\begin{align*}
	\mathbf{c}=\Big(\frac{u_1}{v_1}g(\alpha_1),\ldots,\frac{u_n}{v_n}g(\alpha_n),\eta g_{n-k-1}+(1+\eta S_{\boldsymbol{\alpha}})g_{n-k}\Big),
\end{align*}
thus the Hamming weight 
\begin{align}\label{w}
	\boldsymbol{w}_{\mathbf{c}}\ge n-(n-k)=k.
\end{align}
Note that $\mathcal{C}_{k,n}^{\perp}(\boldsymbol{\alpha},\boldsymbol{v},\eta,\infty)$ is with length $n+1$ and dimension $n+1-k$, by $(\ref{w})$, we complete the proof. $\hfill\Box$
\subsection{The weight distribution of  the $(+)$-ETGRS code}
\begin{theorem}\label{nm}
	Let $A_{n+1-k}$ be the number of codewords in $\mathcal{C}_{k,n}(\boldsymbol{\alpha},\boldsymbol{v},\eta,\infty)$  with weight $n+1-k$, then
	\begin{align*}
	A_{n+1-k}=(q-1)\#N(k,-\eta^{-1},A_{\boldsymbol{\alpha}}).
	\end{align*}
	Furthermore, we have
	
	$(1)$ $\mathcal{C}_{k,n}(\boldsymbol{\alpha},\boldsymbol{v},\eta,\infty)$ is MDS if and only if $\#N(k,-\eta^{-1},A_{\boldsymbol{\alpha}})=0$;
	
	$(2)$ $\mathcal{C}_{k,n}(\boldsymbol{\alpha},\boldsymbol{v},\eta,\infty)$ is NMDS if and only if $\#N(k,-\eta^{-1},A_{\boldsymbol{\alpha}})>0$.
\end{theorem}

{\bf Proof}. Firstly, we show that $\mathcal{C}_{k,n}(\boldsymbol{\alpha},\boldsymbol{v},\eta,\infty)$ is MDS or AMDS.

By the definition, for any $\mathbf{c}_{f}\in\mathcal{C}_{k,n}(\boldsymbol{\alpha},\boldsymbol{v},\eta,\infty)$,  there exists some $$f(x)=f_{k-1}\eta( x^{k}+\eta^{-1}x^{k-1})+\sum\limits_{i=0}^{k-2}f_ix^{i}\in\mathbb{F}_q[x]$$ such that \begin{align*}
 	\mathbf{c}_{f}=(v_1f(\alpha_1),\ldots,v_nf(\alpha_n),f_{k-1}).
 \end{align*}
 
 Let $\boldsymbol{w}_{\mathbf{c}_f}$ be the Hamming weight of $\mathbf{c}_f$, then we have the following two cases. 
 
{\bf Case $1$}. If $f_{k-1}=0$, then $\deg f(x)\le k-2$. Thus
\begin{align*}
	\boldsymbol{w}_{\mathbf{c}_f}\ge n-(k-2)=n-k+2.
\end{align*}

{\bf Case $2$}. If $f_{k-1}\neq0$, then $\deg f(x)= k$. Thus 
\begin{align*}
\boldsymbol{w}_{\mathbf{c}_f}\ge n-k+1.
\end{align*}

Note that $\mathbf{c}_{f}\in\mathcal{C}_{k,n}(\boldsymbol{\alpha},\boldsymbol{v},\eta,\infty)$ is with length $n+1$ and dimension $k$, by Cases $1$-$2$, we know that $\mathcal{C}_{k,n}(\boldsymbol{\alpha},\boldsymbol{v},\eta,\infty)$ is MDS or AMDS.

Next, we determine $A_{n-k+1}$.  
In fact, by Cases $1$-$2$, we know that $$\boldsymbol{w}_{\mathbf{c}_f}= n-k+1 \text{~if and only if~}~\#\{\alpha\in\boldsymbol{A}_{\alpha}~\big|~f(\alpha)=0\}=k,$$
namely, there exists some $k$-element subset $A\subsetneq A_{\boldsymbol{\alpha}}$ and $\lambda\neq 0$, such that 
 	\begin{align}\label{f1}
 	f(x)=\lambda \prod_{\alpha\in A}(x-\alpha)=\lambda\Big( x^k-\sum\limits_{\alpha\in A}\alpha x^{k-1}+\sum\limits_{i=0}^{k-2}(-1)^{k-i}\sum\limits_{I\subsetneq A,|I|=i}\prod_{\alpha\in I}\alpha x^{i}\Big).
 	\end{align}
It implies that \begin{align*}
 	A_{n-k+1}=&\#\bigg(V_{k,1,k-1,\eta}\cap\Big\{f(x)=\lambda \prod_{\alpha\in A}(x-\alpha)~\Big|~\lambda\neq 0,~A\subsetneq A_{\boldsymbol{\alpha}},~|A|=k\Big\}\bigg)\\
 	=&(q-1)\#\Big\{A~\Big|\sum\limits_{\alpha\in A}\alpha=-\eta^{-1},~A\subsetneq A_{\boldsymbol{\alpha}},~|A|=k\Big\}\\
 	=&(q-1)\#N(k,-\eta^{-1},A_{\boldsymbol{\alpha}}).
 \end{align*}
Thus, we have the following two assertions.\\

$(1)$ $\mathcal{C}_{k,n}(\boldsymbol{\alpha},\boldsymbol{v},\eta,\infty)$ is MDS if and only if $\#N(k,-\eta^{-1},A_{\boldsymbol{\alpha}})=0$;

$(2)$ $\mathcal{C}_{k,n}(\boldsymbol{\alpha},\boldsymbol{v},\eta,\infty)$ is AMDS if and only if $\#N(k,-\eta^{-1},A_{\boldsymbol{\alpha}})>0$.\\

By the above discussions, Corollary \ref{c1}, and the fact that the dual code of an MDS code is MDS, we get the desired results.  $\hfill\Box$\\

By Theorem \ref{nm} and Lemma \ref{NMDS}, we get the following theorem directly.
\begin{theorem}\label{tw}
	 The weight distributions of $\mathcal{C}_{k,n}(\boldsymbol{\alpha},\boldsymbol{v},\eta,\infty)$ and $\mathcal{C}_{k,n}^{\perp}(\boldsymbol{\alpha},\boldsymbol{v},\eta,\infty)$ are 
{\small	 
	\begin{align*}
	&A_{n+1-k+s}\\
	=&\begin{cases}
	(q-1)\#N(k,-\eta^{-1},A_{\boldsymbol{\alpha}}),&\text{if}~s=0;\\
	\binom{n+1}{k-s}\sum\limits_{j=0}^{s-1}(-1)^j\binom{n+1-k+s}{j}(q^{s-j}-1)\!+\!(-1)^s	(q-1)\binom{k}{s}\#N(k,-\eta^{-1},A_{\boldsymbol{\alpha}}),&\text{if}~s=1,\dots,k,
	\end{cases}
	\end{align*}
	and
	\begin{align*}
	&A_{k+s}^{\perp}\\
	=&\begin{cases}
	(q-1)\#N(k,-\eta^{-1},A_{\boldsymbol{\alpha}}),&\text{if}~s=0;\\
	\binom{n+1}{k+s}\sum\limits_{j=0}^{s-1}(-1)^j\binom{k+s}{j}(q^{s-j}-1)\!+\!(-1)^s(q-1)\binom{k}{s}\#N(k,-\eta^{-1},A_{\boldsymbol{\alpha}}),&\text{if}~s=1,\dots,n+1-k,
	\end{cases}
	\end{align*}}
respectively.
\end{theorem}

By Lemma \ref{SSS} and Theorem \ref{tw}, we have the following corollaries directly.
\begin{corollary}
	The weight distributions of $\mathcal{C}_{k,q}(\mathbb{F}_q,\boldsymbol{v},\eta,\infty)$ and $\mathcal{C}_{k,q}^{\perp}(\mathbb{F}_q,\boldsymbol{v},\eta,\infty)$ are 
	{\small	 
		\begin{align*}
		&A_{q+1-k+s}\\
		=&\begin{cases}
		\frac{(q-1)}{q}\binom{q}{k},&\!\!\!\text{if}~p\nmid k,\!~s=0;\\
		\binom{q+1}{k-s}\sum\limits_{j=0}^{s-1}(-1)^j\binom{q+1-k+s}{j}(q^{s-j}-1)\!+\!(-1)^s\frac{(q-1)}{q}\binom{k}{s}\binom{q}{k},&\!\!\!\text{if}~p\nmid k,\!~s=1,\dots,k;\\
		\frac{(q-1)}{q}\Big(\!\binom{q}{k}+(-1)^{k+\frac{k}{p}+1}\binom{\frac{q}{p}}{\frac{k}{p}}\!\Big),&\!\!\!\text{if}~p\mid k,\!~s=0;\\
		\binom{q+1}{k-s}\sum\limits_{j=0}^{s-1}(-1)^j\binom{q+1-k+s}{j}(q^{s-j}-1)\!+\!(-1)^s\frac{(q-1)}{q}\binom{k}{s}\Big(\!\binom{q}{k}+(-1)^{k+\frac{k}{p}+1}\binom{\frac{q}{p}}{\frac{k}{p}}\!\Big),&\!\!\!\text{if}~p\mid k,\!~s=1,\dots,k,
		\end{cases}
		\end{align*}
		and
		\begin{align*}
		&A_{k+s}^{\perp}\\
		=&\begin{cases}
		\frac{(q-1)}{q}\binom{q}{k},&\!\!\!\text{if}~p\nmid k,\!~s=0;\\
		\binom{q+1}{k+s}\sum\limits_{j=0}^{s-1}(-1)^j\binom{k+s}{j}(q^{s-j}-1)\!+\!(-1)^s\frac{(q-1)}{q}\binom{k}{s}\binom{q}{k},&\!\!\!\text{if}~p\nmid k,\!~s=1,\dots,q+1-k;\\
		\frac{(q-1)}{q}\Big(\!\binom{q}{k}+(-1)^{k+\frac{k}{p}+1}\binom{\frac{q}{p}}{\frac{k}{p}}\!\Big),&\!\!\!\text{if}~p\mid k,\!~s=0;\\
			\binom{q+1}{k+s}\sum\limits_{j=0}^{s-1}(-1)^j\binom{k+s}{j}(q^{s-j}-1)\!+\!(-1)^s\frac{(q-1)}{q}\binom{k}{s}\Big(\!\binom{q}{k}+(-1)^{k+\frac{k}{p}+1}\binom{\frac{q}{p}}{\frac{k}{p}}\!\Big),&\!\!\!\text{if}~p\mid k,\!~s=1,\dots,q+1-k,
		\end{cases}
		\end{align*}}
	respectively.

\end{corollary}

\begin{corollary}
	If $A_{\boldsymbol{\alpha}}=\mathbb{F}_q^{*}$, then the weight distributions of ${\small \mathcal{C}_{k,q-1}(\boldsymbol{\alpha},\boldsymbol{v},\eta,\infty)}$ and \newline${\small \mathcal{C}_{k,q-1}^{\perp}(\boldsymbol{\alpha},\boldsymbol{v},\eta,\infty)}$ are 
	{\small	 
		\begin{align*}
		&A_{q-k+s}\\
		=&\begin{cases}
		\frac{(q-1)}{q}\Big(\!\binom{q}{k}+(-1)^{k+\lfloor\frac{k}{p}\rfloor+1}\binom{\frac{q}{p}-1}{\lfloor\frac{k}{p}\rfloor}\!\Big),&~\text{if}~s=0;\\
		\binom{q}{k-s}\sum\limits_{j=0}^{s-1}(-1)^j\binom{q-k+s}{j}(q^{s-j}-1)\!+\!(-1)^s\frac{(q-1)}{q}\binom{k}{s}\Big(\!\binom{q}{k}+(-1)^{k+\lfloor\frac{k}{p}\rfloor+1}\binom{\frac{q}{p}-1}{\lfloor\frac{k}{p}\rfloor}\!\Big),&~\text{if}~s=1,\dots,k,
		\end{cases}
		\end{align*}
		and
		\begin{align*}
		&A_{k+s}^{\perp}\\
		=&\begin{cases}
		\frac{(q-1)}{q}\Big(\!\binom{q}{k}+(-1)^{k+\lfloor\frac{k}{p}\rfloor+1}\binom{\frac{q}{p}-1}{\lfloor\frac{k}{p}\rfloor}\!\Big),&~\text{if}~s=0;\\
		\binom{q}{k+s}\sum\limits_{j=0}^{s-1}(-1)^j\binom{k+s}{j}(q^{s-j}-1)\!+\!(-1)^s\frac{(q-1)}{q}\binom{k}{s}\Big(\!\binom{q}{k}++(-1)^{k+\lfloor\frac{k}{p}\rfloor+1}\binom{\frac{q}{p}-1}{\lfloor\frac{k}{p}\rfloor}\!\Big),&~\text{if}~s=1,\dots,q-k,
		\end{cases}
		\end{align*}}
	respectively.
	
\end{corollary}
\subsection{The non-GRS (non-EGRS) property for the $(+)$-ETGRS code}
In this subsection, we show that $(+)$-ETGRS codes are not GRS or EGRS by using the Schur product.

 In the following lemma, we give the Schur square of the $(+)$-ETGRS code.
\begin{lemma}\label{LSP}If $k\ge 3$, then the following two assertions hold. \\
	
	$(1)$ For $k\ge \frac{n+1}{2}$, $$\mathcal{C}_{k,n}^{2}(\boldsymbol{\alpha},\boldsymbol{v},\eta,\infty)=\mathbb{F}_{q}^{n+1}.$$
	
	$(2)$ For $3\le k\le \frac{n}{2}$,
	\begin{align*}	
	&\mathcal{C}_{k,n}^{2}(\boldsymbol{\alpha},\boldsymbol{v},\eta,\infty)\\
	=&\begin{cases}
\mathcal{C}_{2k,n}\big(\boldsymbol{\alpha},\boldsymbol{v}^2,2^{-1}\eta\big),\quad&\text{if~$q$~is odd};\\
\Big\{(v_1^2f({\alpha_{1}}),\ldots,v_n^2f(\alpha_{n}),f_{2k})~|~f(x)=f_{2k}x^{2k}+\sum\limits_{i=0}^{2k-2}f_ix^i\in\mathbb{F}_q[x]\Big\},\quad&\text{if~$q$~is even}.
		\end{cases}
	\end{align*}
	
\end{lemma}

{\bf Proof}. It is enough to prove that $(1)$ and $(2)$ are both true for $\boldsymbol{v}=\mathbf{1}$. In fact, by the definition of $\mathcal{C}_{k,n}(\boldsymbol{\alpha},\boldsymbol{1},\eta,\infty)$ and Remark \ref{r1}, for $k\ge3$, we have
{\small\begin{align*}\begin{aligned}
&\mathcal{C}_{k,n}^{2}(\boldsymbol{\alpha},\boldsymbol{1},\eta,\infty)\\
=&\left\langle(\boldsymbol{\alpha}^{i+j},0),(\boldsymbol{\alpha}^{i}\star(\boldsymbol{\alpha}^{k-1}+\eta\boldsymbol{\alpha}^{k}),0), (\boldsymbol{\alpha}^{k-1}+\eta\boldsymbol{\alpha}^{k},1)^2~(i=0,\ldots,k-2;j=0,\ldots,k-2)\right\rangle\\
=&\left\langle(\boldsymbol{\alpha}^{i},0),(\boldsymbol{\alpha}^{2k-4+j}+\eta\boldsymbol{\alpha}^{2k-3+j},0), (\boldsymbol{\alpha}^{2k-2}+2\eta\boldsymbol{\alpha}^{2k-1}+\eta^2\boldsymbol{\alpha}^{2k},1)~(i=0,\ldots,2k-4;j=0,1)\right\rangle\\
=&\left\langle(\boldsymbol{\alpha}^{i},0),(\boldsymbol{\alpha}^{2k-3}+\eta\boldsymbol{\alpha}^{2k-2},0), (\boldsymbol{\alpha}^{2k-2}+2\eta\boldsymbol{\alpha}^{2k-1}+\eta^2\boldsymbol{\alpha}^{2k},1)~(i=0,\ldots,2k-3)\right\rangle\\
=&\left\langle(\boldsymbol{\alpha}^{i},0), (\boldsymbol{\alpha}^{2k-2}+2\eta\boldsymbol{\alpha}^{2k-1}+\eta^2\boldsymbol{\alpha}^{2k},1)~(i=0,\ldots,2k-2)\right\rangle\\
=&\left\langle(\boldsymbol{\alpha}^{i},0), (2\boldsymbol{\alpha}^{2k-1}+\eta\boldsymbol{\alpha}^{2k},1)~(i=0,\ldots,2k-2)\right\rangle.\\
=&\begin{cases}
\left\langle(\boldsymbol{\alpha}^{i},0), (2\boldsymbol{\alpha}^{2k-1}+\eta\boldsymbol{\alpha}^{2k},1)~(i=0,\ldots,2k-2)\right\rangle,\quad&\text{if~} 2k-2\le n-2;\\
\left\langle(\boldsymbol{\alpha}^{i},0), (\boldsymbol{0},1)~(i=0,\ldots,n-1)\right\rangle,\quad&\text{if~}2k-2\ge n-1;
\end{cases}\\
=&\begin{cases}
\left\langle(\boldsymbol{\alpha}^{i},0), (2\boldsymbol{\alpha}^{2k-1}+\eta\boldsymbol{\alpha}^{2k},1)~(i=0,\ldots,2k-2)\right\rangle,\quad&\text{if~}k\le \frac{n}{2};\\
\mathbb{F}_q^{n+1},\quad&\text{if~} k\ge\frac{n+1}{2}.
\end{cases}
\end{aligned}
\end{align*}}$\hfill\Box$

\begin{theorem}
	For $3\le k\le n-2$, $\mathcal{C}_{k,n}(\boldsymbol{\alpha},\boldsymbol{v},\eta,\infty)$ is not GRS or EGRS.
\end{theorem}

{\bf Proof.} We give  the proof by the following two cases.

{\bf Case $1$}. If $3\le k\le\frac{n}{2}$, then $2k\le n$. Now by Lemma \ref{LSP}, one has
 $$\dim \big(\mathcal{C}_{k,n}^{2}(\boldsymbol{\alpha},\boldsymbol{v},\eta,\infty)\big)=2k.$$ And then by Proposition \ref{pr} and Remark \ref{es}, we know that
 $\mathcal{C}_{k,n}(\boldsymbol{\alpha},\boldsymbol{v},\eta,\infty)$ is not GRS or EGRS.
 
 {\bf Case $2$}. If $n-2\ge k\ge \frac{n}{2}+1$, then $n-k-2\ge 0$ and $2k-n\ge 2$. Now by Theorem $\ref{dt}$, we know that $\boldsymbol{c}_{i}\in \mathcal{C}_{k,n}^{\perp}(\boldsymbol{\alpha},\boldsymbol{v},\eta,\infty)$ $(i=1,2,3)$, where
 \begin{align*}
 & \boldsymbol{c}_1=\Big(~\frac{u_1}{v_1}\alpha_1^{n-k-2},~\frac{u_2}{v_2}\alpha_{2}^{n-k-2},\ldots,\frac{u_n}{v_n}\alpha_{n}^{n-k-2},0~\Big),\\
 &\boldsymbol{c}_2=\Big(~\frac{u_1}{v_1}\alpha_1^{n-k-1},~\frac{u_2}{v_2}\alpha_{2}^{n-k-1},\ldots,\frac{u_n}{v_n}\alpha_{n}^{n-k-1},\eta~\Big),\\
 & \boldsymbol{c}_3=\Big(~\frac{u_1}{v_1}\alpha_1^{n-k},~\frac{u_2}{v_2}\alpha_{2}^{n-k},\ldots,\frac{u_n}{v_n}\alpha_{n}^{n-k},1+\eta S_{\boldsymbol{\alpha}}~\Big).
 \end{align*}Thus
 \begin{align*}
 \boldsymbol{c}=\boldsymbol{c}_1\star\boldsymbol{c}_3-\boldsymbol{c}_2\star\boldsymbol{c}_2=\big(0,0,\ldots,0,\eta^2~\big)\in (\mathcal{C}_{k,n}^{\perp}(\boldsymbol{\alpha},\boldsymbol{v},\eta,\infty))^{2}.
 \end{align*}
 
 For an $[n+1,k]$ GRS (EGRS) code $\mathcal{C}$, by proposition \ref{pr}, we know that $(\mathcal{C}^{\perp})^2$ is an $[n+1,2(n-k)+1]$ GRS  (EGRS)  code, and then the minimum Hamming distance
 \begin{align*}
 	d=(n+1)-(2(n-k)+1)+1=2k-n+1\ge 2.
 \end{align*}
 Thus  $\boldsymbol{c}\notin(\mathcal{C}^{\perp})^2$, and so $\mathcal{C}_{k,n}^{\perp}(\boldsymbol{\alpha},\boldsymbol{v},\eta,\infty)$ is not GRS or EGRS, which implies that  $\mathcal{C}_{k,n}(\boldsymbol{\alpha},\boldsymbol{v},\eta,\infty)$ is not GRS or EGRS.$\hfill\Box$
\section{Self-dual or almost self-dual $(+)$-ETGRS codes}	
\subsection{A sufficient and necessary condition for a $(+)$-ETGRS code to be self-orthogonal}
$\big(\mathcal{C}_{k,n}^{2}(\mathbb{F}_q,\boldsymbol{1},\eta,\infty)\big)^{\perp}$ is given in the following lemma.
 \begin{lemma}\label{LSPd} For $3\le k\le q-2$, we have the following two assertions.\\
 	
 	$(1)$ If $k\ge \frac{q+1}{2}$, then $$\big(\mathcal{C}_{k,n}^{2}(\mathbb{F}_q,\boldsymbol{1},\eta,\infty)\big)^{\perp}=\{\boldsymbol{0}\}.$$
 	
 	$(2)$ If $3\le k\le \frac{q}{2}$, then
 	 
 	 \begin{align*}	
 	  &\big(\mathcal{C}_{k,n}^{2}(\mathbb{F}_q,\boldsymbol{1},\eta,\infty)\big)^{\perp}\\
 	 =&\begin{cases}
 	\Big\{\big(g({\alpha_{1}}),\ldots,g(\alpha_{q}),2^{-1}\eta g_{q-1-2k}+g_{q-2k}\big)~|~\deg g(x)\le q-2k\Big\},\quad&\text{if~$q$~is odd};\\
    \Big\{(g({\alpha_{1}}),\ldots,g(\alpha_{q}),g_{q-1-2k})~|~\deg g(x)\le q-2k\Big\},\quad&\text{if~$q$~is even},
 	 \end{cases}
 	 \end{align*}
 	where $g_{q-1-k}$ and $g_{q-k}$ are the coefficients of $x^{q-1-2k}$ and $x^{q-2k}$ in $g(x)$, respectively.
 \end{lemma}

{\bf Proof}. By  Lemma $\ref{LSP}$, we can obtain $\mathcal{C}_{k,n}^2(\mathbb{F}_q,\boldsymbol{1},\eta,\infty)$.

 For $q$ odd, note that $\mathcal{C}_{k,n}^2(\mathbb{F}_q,\boldsymbol{1},\eta,\infty)$ is a $(+)$-ETGRS code, $\big(\mathcal{C}_{k,n}^2(\mathbb{F}_q,\boldsymbol{1},\eta,\infty)\big)^{\perp}$ can be obtained based on Theorem \ref{dt}.
 
 For $q$ even, by $(\ref{lq1})$ we can verify that $(2)$ is true directly.$\hfill\Box$\\

 
 Basing on Lemma \ref{lso}, Remark \ref{r0} and Lemma \ref{LSPd}, we can get the following Theorems \ref{tso1}-\ref{tso2} directly.
 \begin{theorem}\label{tso1} For $3\le k\le \frac{q}{2}$,  we have the following two assertions.
	
	$(1)$ If $q$ is odd, then $\mathcal{C}_{k,n}(\boldsymbol{\alpha},\boldsymbol{v},\eta)$ is  self-orthogonal if and only if  there exists some 
	$g(x)=\sum\limits_{i=0}^{q-2k}g_{i}x^{i}\in\mathbb{F}_q[x]$ such that
	\begin{align}\label{du1}
	\eta g_{q-1-2k}+2g_{q-2k}=0,~~ g(\alpha_j)=v_j^2~(j=1,\ldots,n),~~g(\beta)=0~(\forall\beta\in\mathbb{F}_q\backslash A_{\boldsymbol{\alpha}}). 
	\end{align}
	
	$(2)$ If $q$ is even, $\mathcal{C}_{k,n}(\boldsymbol{\alpha},\boldsymbol{v},\eta)$ is  self-orthogonal if and only if  there exists some 
	$g(x)=\sum\limits_{i=0}^{q-2k}g_{i}x^{i}\in\mathbb{F}_q[x]$ such that\begin{align}\label{du2}
	g_{q-1-2k}=0,~~ g(\alpha_j)^{\frac{q}{2}}=v_j~(j=1,\ldots,n),~~g(\beta)=0~(\forall\beta\in\mathbb{F}_q\backslash A_{\boldsymbol{\alpha}}). 
	\end{align}
	
\end{theorem}

By Theorem \ref{tso1}, we have the following corollary.
\begin{corollary}\label{tso12} For $3\le k\le \frac{q}{2}$ and $l\in\{3,\ldots,k\}$, if $\mathcal{C}_{k,n}(\boldsymbol{\alpha},\boldsymbol{v},\eta)$ is  self-orthogonal, then $\mathcal{C}_{l,n}(\boldsymbol{\alpha},\boldsymbol{v},\eta)$ is  self-orthogonal.	
\end{corollary}

{\bf Proof}. If $\mathcal{C}_{k,n}(\boldsymbol{\alpha},\boldsymbol{v},\eta)$ is  self-orthogonal, then there exists some $g(x)=\sum\limits_{i=0}^{q-2k}g_{i}x^{i}\in\mathbb{F}_q[x]$ such that $(\ref{du1})$ or $(\ref{du2})$ holds. 
For $l\in\{3,\ldots,k-1\}$, let $$h(x)=\sum_{i=q-2k+1}^{q-2l}g_ix^{i}+g(x)~\text{with}~g_i=0~(i=q-2k+1,\ldots,q-2l),$$ then we can check that $h(x)$ satisfies  $(\ref{du1})$ or $(\ref{du2})$, thus $\mathcal{C}_{l,n}(\boldsymbol{\alpha},\boldsymbol{v},\eta)$ is  self-orthogonal. $\hfill\Box$\\

Especially,  we can get a sufficient and necessary condition for  $\mathcal{C}_{k,n}(\boldsymbol{\alpha},\boldsymbol{v},\eta)$ to be self-dual.
\begin{corollary}\label{tsol0}
	For $3\le k\le \frac{q}{2}$, if $u_j=-\prod\limits_{i=1,i\neq j}^{2k}(\alpha_j-\alpha_i)^{-1}$ $(j=1,\ldots,2k)$, then we have the following two assertions.

	$(1)$ If $q$ is odd, then ${\mathcal{C}}_{k,2k}(\boldsymbol{\alpha},\boldsymbol{v},\eta)$ is self-dual if and only if there exists some  $\lambda\in\mathbb{F}_q^{*}$ such that
	\begin{align*}
	\eta S_{\boldsymbol{\alpha}}+2= 0~~~~\text{and}~~\lambda u_j=v_j^2~(j=1,\ldots,n). 
	\end{align*}
	
	$(2)$ If $q$ is even, 
	then ${\mathcal{C}}_{k,2k}(\boldsymbol{\alpha},\boldsymbol{v},\eta)$ is self-dual if and only if there exists some  $\lambda\in\mathbb{F}_q^{*}$ such that
	\begin{align*}
	S_{\boldsymbol{\alpha}}=0~~~~\text{and}~~\lambda u_j^{\frac{q}{2}}=v_j~(j=1,\ldots,2k). 
	\end{align*}
\end{corollary}

{\bf Proof.} By Theorem \ref{tso1}, we know that  $\mathcal{C}_{k,n}(\boldsymbol{\alpha},\boldsymbol{v},\eta)$ is  self-dual if and only if  there exists some $\lambda\in\mathbb{F}_q^{*}$ such that $g(x)=\lambda\prod\limits_{\beta\in\mathbb{F}_q\backslash A_{\boldsymbol{\alpha}}}(x-\beta)$ satisfies $(\ref{du1})$ or $(\ref{du2})$. Now by $$\lambda\prod_{\beta\in\mathbb{F}_q\backslash A}(\alpha_j-\beta)=-\lambda\prod\limits_{i=1,i\neq j}^{2k}(\alpha_j-\alpha_i)^{-1}~ (j=1,\ldots,2k),$$  we get the desired results directly.$\hfill\Box$\\

\begin{theorem}\label{tso2} For $3\le k\le \frac{q}{2}$,  we have the following two assertions.
	
	$(1)$ If $q$ is odd, then ${\mathcal{C}}_{k,n}(\boldsymbol{\alpha},\boldsymbol{v},\eta,\infty)$ is  self-orthogonal if and only if  there exists some
	$g(x)=\sum\limits_{i=0}^{q-2k}g_{i}x^{i}\in\mathbb{F}_q[x]$ such that
	\begin{align}\label{ddu1}
	2^{-1}\eta g_{q-1-2k}+g_{q-2k}=1,~~ g(\alpha_j)=v_j^2~(j=1,\ldots,n),~~g(\beta)=0~(\forall\beta\in\mathbb{F}_q\backslash A_{\boldsymbol{\alpha}}). 
	\end{align}
	
	$(2)$ If $q$ is even, $\mathcal{C}_{k,n}(\boldsymbol{\alpha},\boldsymbol{v},\eta,\infty)$ is  self-orthogonal if and only if  there exists some
	$g(x)=\sum\limits_{i=0}^{q-2k}g_{i}x^{i}\in\mathbb{F}_q[x]$ such that\begin{align}\label{ddu2}
	g_{q-1-2k}=1,~~ g(\alpha_j)^{\frac{q}{2}}=v_j~(j=1,\ldots,n),~~g(\beta)=0~(\forall\beta\in\mathbb{F}_q\backslash A_{\boldsymbol{\alpha}}). 
	\end{align}
\end{theorem}

By Theorem \ref{tso2}, we have the following corollary.
\begin{corollary}\label{tso13} For $3\le k\le \frac{q}{2}$ and $l\in\{3,\ldots,k\}$, if $0\notin A_{\boldsymbol{\alpha}}$ and $\mathcal{C}_{k,n}(\boldsymbol{\alpha},\boldsymbol{v},\eta,\infty)$ is  self-orthogonal, then $\mathcal{C}_{l,n}(\boldsymbol{\alpha},\boldsymbol{\alpha}^{k-l}\star\boldsymbol{v},\eta,\infty)$ is  self-orthogonal.
\end{corollary}

{\bf Proof}. If $\mathcal{C}_{k,n}(\boldsymbol{\alpha},\boldsymbol{v},\eta,\infty)$ is  self-orthogonal, then there exists some $g(x)=\sum\limits_{i=0}^{q-2k}g_{i}x^{i}\in\mathbb{F}_q[x]$ such that $(\ref{ddu1})$ or $(\ref{ddu2})$ holds. For $0\notin A_{\boldsymbol{\alpha}}$ and $l\in\{3,\ldots,k-1\}$, let $$h(x)=x^{2(k-l)}g(x)=\sum_{i=0}^{q-2l}\bar{g}_{i}x^{i},$$ then for $q$ odd, we have
\begin{align}\label{4.5}
\eta \bar{g}_{q-1-2l}+2\bar{g}_{q-2l}=0,~~ h(\alpha_j)=\alpha_j^{2(k-l)}v_j^2~(j=1,\ldots,n),~~h(\beta)=0~(\forall\beta\in\mathbb{F}_q\backslash A_{\boldsymbol{\alpha}}). 
\end{align}
For $q$ even, we have 
\begin{align}\label{4.6}
\bar{g}_{q-1-2l}=0,~~ h(\alpha_j)^{\frac{q}{2}}=\alpha_j^{k-l}v_j~(j=1,\ldots,n),~~h(\beta)=0~(\forall\beta\in\mathbb{F}_q\backslash A_{\boldsymbol{\alpha}}). 
\end{align}
Now by Theorem \ref{tso2} and $(\ref{4.5})$-$(\ref{4.6})$,  $\mathcal{C}_{l,n}(\boldsymbol{\alpha},\boldsymbol{\alpha}^{k-l}\star\boldsymbol{v},\eta,\infty)$ is  self-orthogonal. $\hfill\Box$\\

By Theorem \ref{tso2}, if ${\mathcal{C}}_{k,n}(\boldsymbol{\alpha},\boldsymbol{v},\eta,\infty)$ is  self-orthogonal, then $q-2k\ge q-n$, namely, $n\ge 2k$. Note that the length of ${\mathcal{C}}_{k,n}(\boldsymbol{\alpha},\boldsymbol{v},\eta,\infty)$ is $n+1$, thus we have the following corollary.
\begin{corollary}
	There is no any self-dual $(+)$-ETGRS code.\\
\end{corollary}

By Theorem \ref{tso2}, in the similar proof as that for Corollary \ref{tsol0}, we have the following corollary.
\begin{corollary}\label{tso11}
	For $3\le k\le \frac{q}{2}$, let $u_j=-\prod\limits_{i=1,i\neq j}^{2k}(\alpha_j-\alpha_i)^{-1}$ $(j=1,\ldots,2k)$, we have the following two assertions.
	
	$(1)$ If $q$ is odd, then ${\mathcal{C}}_{k,2k}(\boldsymbol{\alpha},\boldsymbol{v},\eta,\infty)$ is almost self-dual if and only if there exists some  $\lambda\in\mathbb{F}_q^{*}$ such that
	\begin{align*}
	\lambda (2^{-1}\eta S_{\boldsymbol{\alpha}}+1)=1,~~\text{and}~~ \lambda u_j=v_j^2~(j=1,\ldots,n). 
	\end{align*}
	
	$(2)$ If $q$ is even, 
	then ${\mathcal{C}}_{k,2k}(\boldsymbol{\alpha},\boldsymbol{v},\eta,\infty)$ is almost self-dual if and only if there exists some  $\lambda\in\mathbb{F}_q^{*}$ such that
	\begin{align*}
	\lambda S_{\boldsymbol{\alpha}}=1,~~\text{and}~~ (\lambda u_j)^{\frac{q}{2}}=v_j~(j=1,\ldots,n). 
	\end{align*}
	
\end{corollary}

 

\subsection{The Construction for the self-orthogonal $(+)$-TGRS (ETGRS) code}
\subsubsection{The case for  $q$ even}
For any  $b\in\mathbb{F}_q^{*}$ and $3\le t\le q-4$, by Remark \ref{rsss}, we know that
$$N(t,b,\mathbb{F}_q^{*})=\big\{\{x_1,\ldots,x_{2k}\}\subseteq D\big|x_1+\cdots+x_{t}=b\big\}\neq \emptyset,$$and
$$N(t,b,\mathbb{F}_q)=\big\{\{x_1,\ldots,x_{2k}\}\subseteq D\big|x_1+\cdots+x_{t}=b\big\}\neq \emptyset.$$ Now by Corollaries $\ref{tsol0}$ and $\ref{tso11}$, we can obtain the following theorem directly.
\begin{theorem}\label{CDA}
	For $q$ even and $3\le k\le \frac{q-2}{2}$, let $\eta\in\mathbb{F}_q^{*}$, $\boldsymbol{\alpha}=(\alpha_1,\ldots,\alpha_{2k})$, and  $$\boldsymbol{v}=(v_1,\ldots,v_{2k})\text{~ with~} v_j=\prod\limits_{i=1,i\neq j}^{2k}(\alpha_j-\alpha_i)^{-\frac{q}{2}}~ (j=1,\ldots,2k).$$ Then 
	
	$(1)$ for $A_{\boldsymbol{\alpha}}\in N(2k,0,\mathbb{F}_q)$,  ${\mathcal{C}}_{k,2k}(\boldsymbol{\alpha},\boldsymbol{v},\eta)$ is self-dual; 
	
	$(2)$ for $A_{\boldsymbol{\alpha}}\in N(2k,1,\mathbb{F}_q^{*})$,  ${\mathcal{C}}_{k,2k}(\boldsymbol{\alpha},\boldsymbol{v},\eta,\infty)$ is almost self-dual. 
\end{theorem}

By  Theorem \ref{CDA}, Corollaries $\ref{tso12}$ and $\ref{tso13}$, we have the following corollary.
\begin{corollary} Let  $q$, $k$, $\eta$, $\boldsymbol{\alpha}$ and $\boldsymbol{v}$ be given in Theorem \ref{CDA}.
	If $l\in\mathbb{Z}^{+}$ with $3\le l\le k-1$, then  
	
	$(1)$ for $A_{\boldsymbol{\alpha}}\in N(2k,0,\mathbb{F}_q)$,  ${\mathcal{C}}_{l,2k}(\boldsymbol{\alpha},\boldsymbol{v},\eta)$ is self-orthogonal;
	
	$(2)$ for $A_{\boldsymbol{\alpha}}\in N(2k,1,\mathbb{F}_q^{*})$,  ${\mathcal{C}}_{l,2k}(\boldsymbol{\alpha},\boldsymbol{\alpha}^{k-l}\star\boldsymbol{v},\eta,\infty)$ is self-orthogonal.
\end{corollary}


\subsubsection{The case for  $q$ odd}
Note that any element in $\mathbb{F}_{p^m}$ is a square element in $\mathbb{F}_{p^{2m}}$, then by Theorem \ref{nm}, Corollaries $\ref{tsol0}$ and  $\ref{tso11}$, we have the following theorem.
\begin{theorem}\label{PCD1}
	For any positive integer $m$ and odd prime $p$, let $3\le k\le \frac{p^{m}-1}{2}$, $q=p^{2m}$ and $\mathrm{F}_{p^{m}}^{*}=\langle\gamma\rangle$, $\boldsymbol{\alpha}=(\alpha_1,\ldots,\alpha_{2k})$ and  $$\boldsymbol{v}=(v_1,\ldots,v_{2k})\text{~ with~} v_j=\Big(-\prod\limits_{i=1,i\neq j}^{2k}(\alpha_j-\alpha_i)^{-1}\Big)^{\frac{1}{2}}~ (j=1,\ldots,2k).$$ Then the following two assertions hold.\\
	
	$(1)$ If $\boldsymbol{\alpha}=(\gamma,\ldots,\gamma^{i_0-1},0,\gamma^{i_0+1},\ldots,\gamma^{k},-\gamma,\ldots,-\gamma^{k})$	and $\eta=2\gamma^{-i_0}$, then ${\mathcal{C}}_{k,2k}(\boldsymbol{\alpha},\boldsymbol{v},\eta)$ is self-dual.\\
	
	$(2)$ If $\boldsymbol{\alpha}=(\gamma,\ldots,\gamma^{k},-\gamma,\ldots,-\gamma^{k}),$ then\\	
	
	$(1.1)$ for $\eta\in\mathbb{F}_{q}\backslash\mathbb{F}_{p^m}$, ${\mathcal{C}}_{k,2k}(\boldsymbol{\alpha},\boldsymbol{v},\eta,\infty)$ is an almost self-dual MDS code;
	
	$(1.2)$ for $N(k,-\eta^{-1},A_{\boldsymbol{\alpha}})>0$,  ${\mathcal{C}}_{k,2k}(\boldsymbol{\alpha},\boldsymbol{v},\eta,\infty)$ is an almost self-dual NMDS code.  \\
	
\end{theorem}

By Theorem \ref{PCD1}, Corollaries $\ref{tso12}$ and $\ref{tso13}$, we have the following corollary.
\begin{corollary}
	Let $q$, $k$, $m$, $\eta$, $\boldsymbol{\alpha}$ and $\boldsymbol{v}$ be given in Theorem \ref{PCD1}. For any integer $l$ with $3\le l\le k$, we have the following two assertions.
	
	$(1)$ If $\boldsymbol{\alpha}=(\gamma,\ldots,\gamma^{i_0-1},0,\gamma^{i_0+1},\ldots,\gamma^{k},-\gamma,\ldots,-\gamma^{k})$	and $\eta=2\gamma^{-i_0}$, then ${\mathcal{C}}_{l,2k}(\boldsymbol{\alpha},\boldsymbol{v},\eta)$ is self-orthogonal.\\
	
	$(2)$ If $\boldsymbol{\alpha}=(\gamma,\ldots,\gamma^{k},-\gamma,\ldots,-\gamma^{k}),$ then \\	
	
	$(1.1)$ for $\eta\in\mathbb{F}_{q}\backslash\mathbb{F}_{p^m}$, ${\mathcal{C}}_{l,2k}(\boldsymbol{\alpha},\boldsymbol{\alpha}^{k-l}\star\boldsymbol{v},\eta,\infty)$ is a self-orthogonal MDS code;
	
	$(1.2)$ for $N(k,-\eta^{-1},A_{\boldsymbol{\alpha}})>0$,  ${\mathcal{C}}_{l,2k}(\boldsymbol{\alpha},\boldsymbol{\alpha}^{k-l}\star\boldsymbol{v},\eta,\infty)$ is a  self-orthogonal NMDS code.  \\
\end{corollary}

Now we give a construction for almost self-dual $(+)$-ETGRS codes by using the trace map. For integers $r$ and $m$ with $r\mid m$, the trace map  from $\mathbb{F}_{p^m}$ to $\mathbb{F}_{p^r}$ is defined as
$$\mathrm{Tr}_r^{m}(x)=x^{p^{m-r}}+x^{p^{m-2r}}+\cdots+x~(\forall x\in\mathbb{F}_{p^m}).$$
Denote $\mathrm{Ker}(\mathrm{Tr}_r^{m})=\{x\in\mathbb{F}_{p^m}|\mathrm{Tr}_r^{m}(x)=0\}$. Since the trace map is uniform, we have 
\begin{align}\label{t1}
	|\mathrm{Ker}(\mathrm{Tr}_r^{m})|=p^{m-r},
\end{align}
and then
\begin{align}\label{t2}
	\mathrm{Tr}_r^{m}(x)=\prod_{\alpha\in\mathrm{Ker}(\mathrm{Tr}_r^{m})}(x-\alpha).
\end{align}
 
Note that any element in $\mathbb{F}_{p^r}$ is a square element in $\mathbb{F}_{p^m}$, by (\ref{t1})-(\ref{t2}), based on Theorem $\ref{tso2}$, we get the following theorem directly.

\begin{theorem}\label{CT}
	For any odd prime $p$,  integers $r$ and $m$ with $2\mid \frac{m}{r}$, let $q=p^m$ and $3\le k\le \frac{q-2}{2}$. If $\eta\in\mathbb{F}_{q}^{*}$, $\boldsymbol{\alpha}=(\alpha_1,\ldots,\alpha_{p^m-p^{m-r}})$ with $A_{\boldsymbol{\alpha}}=\mathbb{F}_{p^m}\backslash \mathrm{Ker}(\mathrm{Tr}_r^{m})$, and   $$\boldsymbol{v}=(v_1,\ldots,v_{j})\text{~ with~} v_j=\mathrm{Tr}_r^{m}(\alpha_j)^{\frac{1}{2}}~(j=1,\ldots,p^m-p^{m-r}).$$ Then
	${\mathcal{C}}_{\frac{p^m-p^{m-r}}{2},p^m-p^{m-r}}(\boldsymbol{\alpha},\boldsymbol{v},\eta,\infty)$ is almost self-dual.
\end{theorem}

By Theorem \ref{CT},  Corollaries \ref{tso12} and \ref{tso13}, we have the following corollary.
\begin{corollary}
 Let $q$, $r$, $m$, $\eta$, $\boldsymbol{\alpha}$ and $\boldsymbol{v}$ be given in Theorem \ref{CT}. Then for any integer $l$ with $3\le l\le \frac{p^m-p^{m-r}}{2}-1$, 
  ${\mathcal{C}}_{l,p^m-p^{m-r}}(\boldsymbol{\alpha},\boldsymbol{\alpha}^{\frac{p^m-p^{m-r}-2l}{2}}\star\boldsymbol{v},\eta,\infty)$ is self-orthogonal.
\end{corollary}

\section{Conclusions }

In this paper, we have the following main results.\\

$(1)$  The parity check  matrix for the $(+)$-ETGRS code is given.\\

$(2)$ The $(+)$-ETGRS code is MDS or NMDS.\\

$(3)$ The $(+)$-ETGRS code is not GRS or EGRS.\\

$(4)$ The weight distribution of the $(+)$-ETGRS code is determined.\\

$(5)$ A  sufficient and necessary condition for any punctured code of the $(+)$-ETGRS code to be self-orthogonal is presented.\\

$(6)$ Several classes of (almost) self-dual MDS or NMDS codes are constructed. \\


\end{document}